\documentclass[fleqn,twoside]{article}
\usepackage{espcrc2}

\usepackage{graphicx}
\usepackage[figuresright]{rotating}

\title{The BlueGene/L Supercomputer\thanks{Talk presented by D.~Chen.}}

\author{Gyan Bhanot\address{IBM T.J. Watson Research Center,
	    Yorktown Heights, NY 10598, USA},
        Dong Chen\addressmark, Alan Gara\addressmark\ 
	and Pavlos Vranas\addressmark} 

\begin{document}

\begin{abstract}
  The architecture of the BlueGene/L massively parallel supercomputer is
  described.  Each computing node consists of a single compute ASIC plus
  256 MB of external memory.  The compute ASIC integrates two 700 MHz 
  PowerPC 440 integer CPU cores, two 2.8 Gflops floating point units,
  4 MB of embedded DRAM as cache, a memory controller for external memory,
  six 1.4 Gbit/s bi-directional ports for a 3-dimensional
  torus network connection, three 2.8 Gbit/s bi-directional ports for
  connecting to a global tree network and a Gigabit Ethernet for I/O.
  65,536 of such nodes are connected into a
  3-d torus with a geometry of 32$\times$32$\times$64.
  The total peak performance of the system is 360 Teraflops and the
  total amount of memory is 16 TeraBytes.
\vspace{1pc}
\end{abstract}

\maketitle

\section{INTRODUCTION}
IBM has previously announced a multi-year initiative to build a petaflops
scale supercomputer for computational biology research \cite{bluegene}. 
The BlueGene/L machine is a first step in this program \cite{bluegeneweb}.
It is based on a different and more general architecture than the original
BlueGene announcement.  In particular, BlueGene/L uses embedded 
PowerPC processor cores developed by IBM Microelectronics \cite{imd}
for ASIC products. 

The lattice community has seen great success of some special-purpose, 
massively parallel machines dedicated for QCD, 
for example, the combined 1 Teraflops QCDSP machine \cite{qcdsp99}
and the follow-on 20 Teraflops QCDOC machine \cite{qcdoc01}
currently being developed at Columbia University in collaboration
with IBM Research.

The BlueGene/L design philosophy has been influenced by QCDSP and QCDOC.
Contrasting to the current commercial approach of building large-scale
supercomputers by clustering general purpose yet complex SMP systems,
BG/L leverages IBM's system-on-a-chip silicon technology and builds
a large parallel system consisting of more than 65,000 nodes,
yet at a significantly lower price/performance and 
power consumption/performance versus conventional approaches.
Compared to QCDOC, BG/L will be using a newer generation of IBM's silicon
technology, with enhancements in single node performance as well as
a more general network supporting hardware point-to-point message passing
(cut-through routing).  This makes BG/L suitable for a wide
variety of applications. 

The BlueGene/L project is a jointly funded research partnership
between IBM and the Lawrence Livermore National Laboratory (LLNL)
as part of the US Department of Energy ASCI Advanced Architecture 
Research Program.  The main research and development effort is
centered at the IBM T.J.~Watson Research Center, with supports from
IBM Enterprise Server Group and IBM Microelectronics.
Application performance and scaling studies have been initiated
with partners at a number of academic and government institutions,
including the San Diego Supercomputing Center and 
the California Institute of Technology.

A large machine with a peak performance of 360 Teraflops is anticipated to
be built at the LLNL in the 2004-2005 time frame.  A smaller, 100
Teraflops machine is expected to be built at the IBM T.J.~Watson Research
Center for computational biology studies.

\section{BG/L OVERVIEW}

\begin{figure*}[thb]
\vspace{-3pc}
\includegraphics[angle=270,scale=0.5,width=\hsize]{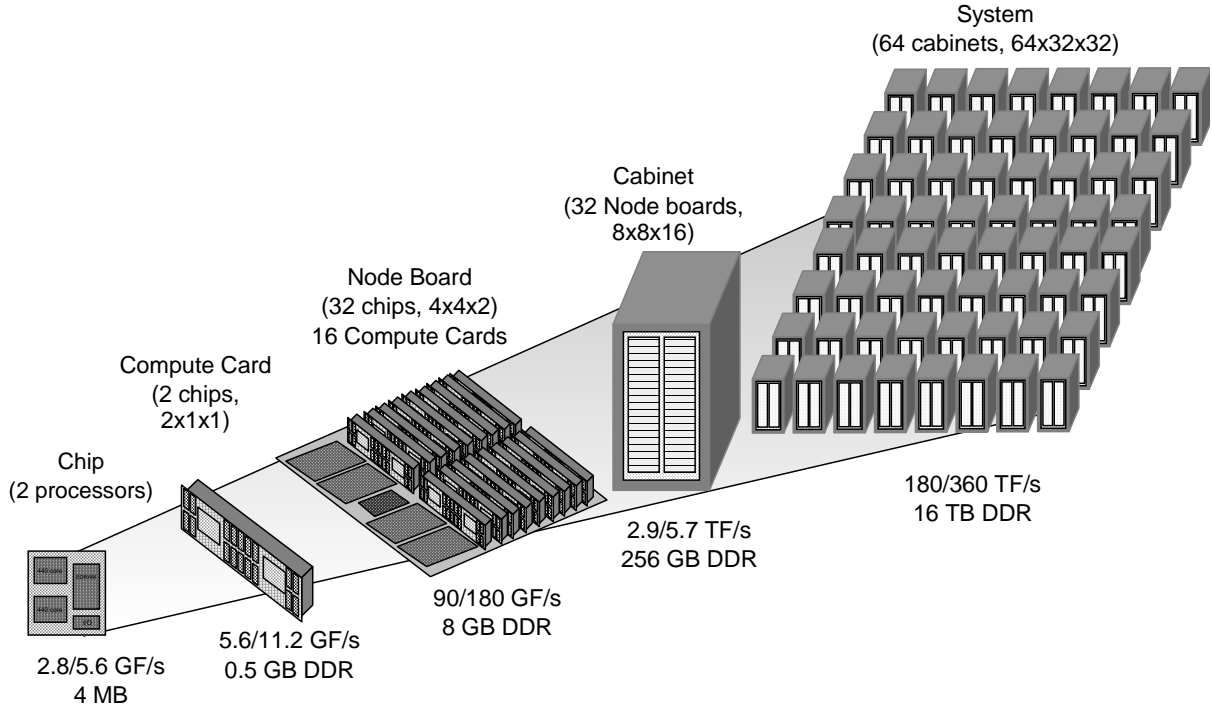}
\vspace{-5pc}
\caption{Building BlueGene/L.  A node ASIC contains 2 CPUs.  2 node ASICs
along with their associated local external memory are built onto a compute
card.  A node board contains 16 compute cards.  32 node boards are then
plugged into both sides of 2 mid-plane boards in a cabinet.  A large
BG/L system contains 64 cabinets, forming a 32$\times$32$\times$64 torus
of compute nodes, with a peak performance of 180/360 Tflops.}
\label{fig:bglbuild}
\end{figure*}

BlueGene/L is a massively parallel, scalable system.  A single parallel
job can use up to 65,536 compute nodes.  The system is configured as a
32$\times$32$\times$64 three-dimensional torus of compute nodes.  Each node
consists of a single compute ASIC and memory.  Each node can support up to
2 GB of local memory.  Balancing cost and application requirements,
our current plan calls for 256 MB of DDR-SDRAM per node.

The ASIC will be manufactured on IBM CMOS CU-11 0.13 micron copper
technology with an expected 11.1 $mm^2$ die size.  This is the
next generation IBM CMOS technology compared
to the one used by QCDOC.  There are two PowerPC 440 32 bit integer CPU
cores in each BG/L ASIC, each core connects to a ``double" floating point unit
capable of 2 fused floating point multiply-adds per CPU clock cycle.
At a target frequency of 700 MHz, each
core can achieve a peak performance of 2.8 Gflops.  In normal operations,
we expect that one CPU will be mainly doing computation while the other one
will be busy handling communications.  However, for certain kinds of
applications, if the communication requirement are small compared to
the amount of compute, or if there are separate computation 
and communication steps, then both cores can be utilized for compute,
leading to 5.6 Gflops peak performance per node.  We therefore quote
the performance per node as 2.8/5.6 Gflops.

Figure \ref{fig:bglbuild} shows the steps of building the BlueGene/L
supercomputer.  2 compute ASIC chips along with their associated local
memory are put onto a compute card.  A node board will have 16 compute cards.
A cabinet includes 2 mid-planes. A total of 32 node boards are
plugged into both sides of the 2 mid-planes.
Within a cabinet, the compute nodes form a geometry of
8$\times$8$\times$16 (8$\times$8$\times$8 per mid-plane) with a
peak performance of 2.9/5.7 Tflops and a total of 256 GB memory.
The BlueGene/L system consists of 64 cabinets connected as a
32$\times$32$\times$64 torus.  The total peak performance is 180/360 Tflops
and the total amount of memory is 16 TB.
The system will occupy an area of about 250 $m^2$, and the total
power is estimated at approximately 1 MW.

In addition to the computing nodes in the system, there are also I/O nodes.
Each I/O node contains the same ASIC as in a compute node, but with
512 MB of memory.  These additional I/O cards are plugged into the
same node board with compute cards.  An I/O node connects to a 
number of compute nodes through a custom high speed network,
and to outside host and disk farm through a Gigabit Ethernet.
These I/O nodes are used to offload the work required for disk I/O
and host communications from the compute nodes.  We plan to install
one I/O node per 64 compute nodes.  The maximum ratio is one I/O node
for every 8 compute nodes.

Besides the active 64 cabinets, the large BG/L system also includes
a number of spare cabinets, Gigabit Ethernet switches racks,
disk I/O racks plus a host computer.  The Gigabit switches connect
to the BG/L I/O nodes, the host computer and the disk farm.

\section{BG/L ARCHITECTURE}

\begin{figure*}[th]
\vspace{-4pc}
\includegraphics[angle=270,scale=0.5,width=\hsize]{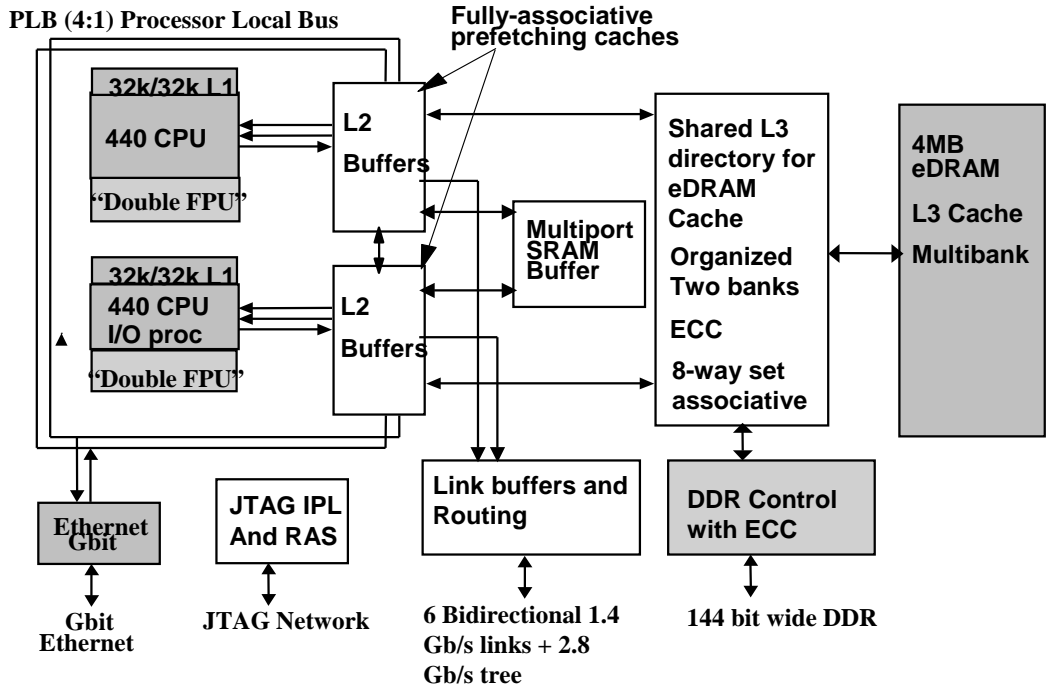}
\vspace{-5pc}
\caption{BlueGene/L node ASIC.  Each ASIC integrates two 32 bit PowerPC 440
integer CPU cores, 2 ``double" floating point FP2 cores, L2 cache buffers, 
4 MB EDRAM as L3 cache, DDR-SDRAM controller for external memory,
high speed torus network logic, global tree network logic,
Gigabit Ethernet and JTAG control interface.}
\label{fig:bglasic}
\end{figure*}

The architecture of the BG/L node ASIC is shown in Figure \ref{fig:bglasic}.
Each ASIC integrates two PowerPC 440 cores, each with a PowerPC 440 FP2 core,
2 small L2 buffers, 4 MB embedded DRAM configured as L3 cache,
DDR-SDRAM memory controller for connecting to external memory,
custom designed high speed torus and tree network logic,
a Gigabit Ethernet and a JTAG control interface.

\subsection{PowerPC core}
The 440 is a standard 32-bit integer microprocessor core product from
IBM Microelectronics.  The same core for a previous generation
technology is used in the QCDOC project.  Each core has 32 general purpose
registers.  It has integrated 32 KB instruction L1 cache and 32 KB data 
L1 cache.  They provide one cycle access from the CPU.
There are three 128 bit buses coming out of the core,
one each dedicated for data read, data write and instruction load.
This core supports all common PowerPC instructions as well as instructions
defined in the PowerPC Book E standard for embedded processors.

The 440 FP2 core is an enhanced ``double" 64-bit floating-point unit.
It consists of a primary and a secondary unit, each of which is a complete
FPU with their own register sets.  The primary FPU supports standard 
PowerPC floating point instructions.  It can do a single fused 64 bit
multiply-add in one processor cycle. 
Through a SIMD like instruction extension, 
both FPU units can be utilized to execute 2 fused 64 bit multiply-adds
per cycle.  In addition, a separate floating point load/store operation
can be executed in parallel to the multiply-adds.  The load/store unit
supports 128 bit ``quad-word" load/store from either the L1 cache 
or the 440 memory interface, to a pair of registers,
one each from the two FP units. 
This increased load/store bandwidth
to FP registers is to match the increased floating point performance.
It is also used to efficiently move data to and from the custom
high-speed network interfaces, as the 32 bit integer unit does not
have the adequate bandwidth to its internal registers to support
high speed data movements.
The floating point instruction extension is also powerful enough to allow the
exchange of register contents between the two FP units while they
are executing multiply-adds, without extra instruction overheads.

\subsection{Memory Subsystem}
Each of the 440 processor core is directly connected to a small 2KB L2 cache,
then to a shared L3 directory which controls 4 MB of
embedded DRAM as the L3 cache.  The L3 controller directly connects to
a DDR-SDRAM controller for external memory.
Both the 4 MB L3 cache and the external DDR-SDRAM are ECC protected.

Because the 440 hardware does not support SMP protocols for its L1 cache, 
the L1 cache of the two processors are not coherent.  We have therefore
implemented a lock box and a small fast multi-port SRAM behind the
two L2 caches to facilitate processor-to-processor communications.
The L2 and L3 caches are coherent for both processors.

A data prefetch engine is built into each L2 cache to reduce the
latency for sequential data accesses from L3 and external memory.
The memory subsystem is being designed for low latency, high bandwidth
accesses to cache and memory.  An L2 hit returns in 6 to 10 processors
cycles, an L3 hit in about 25 cycles, and an L3 miss (loading from external
DRAM) in about 75 cycles.  Various memory bandwidth numbers are listed
in Table \ref{tab:bglqcdoc} as they are compared to QCDOC.

The L2 caches are also directly connected to a set of FIFOs to
allow for fast and high bandwidth access to both a 3-d
torus network and a global tree network.

\subsection{3-D Torus Network}
The 3-d torus is a high speed network used for general-purpose,
point-to-point message passing operations.  Each ASIC has 6 
torus links build in.  On a compute node, these 6 links are connected to 
its 6 nearest neighbors, in the $+x$, $-x$, $+y$, $-y$, $+z$ and $-z$
directions, respectively.  There is one link between a pair of
nearest neighbors.  Each link is a bi-directional serial connection 
with a target speed of 1.4 Gbit/s per direction.

\begin{figure}[thb]
\includegraphics[angle=270,scale=0.5,width=\hsize]{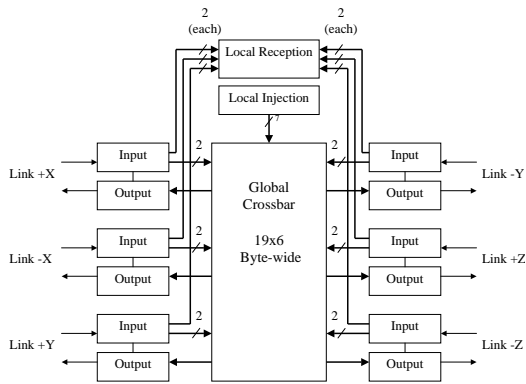}
\vspace{-3pc}
\caption{BlueGene/L torus logic within a node.  2 sets of injections FIFOs
and 2 sets of reception FIFOs are directly connected to two L2 buffers.
6 bi-directional serial links connect to the 6 nearest 
neighbors of the node on a 3-d torus.  A global cross-bar connects all
injection FIFOs, reception FIFOs and 6 links.}
\label{fig:bgltorusarch}
\end{figure}
Figure \ref{fig:bgltorusarch} illustrates the torus logic. 
Within a node ASIC, 2 sets of injection FIFOs and 2 sets of reception
FIFOs are directly connected to the L2 cache.  These FIFOs along with
6 input links and 6 output links are then connected through a
global cross-bar switch.  A packet injected on one node will route
through the 3-d torus network in hardware without any software overhead
until it reaches its destination, where it will then be pulled out by a CPU
from a reception FIFO.  A packet could vary in size from 32 bytes
to 256 bytes in 32-byte granularity.  Even though in the normal operating
mode, we envision that one CPU will be used for compute and 
the other for network and I/O traffic, there is no hardware limitation
as to which CPU can access which FIFO.  The design is symmetric in this
respect.

\begin{figure}[bht]
\vspace{-2pc}
\includegraphics[angle=270,scale=0.5,width=\hsize]{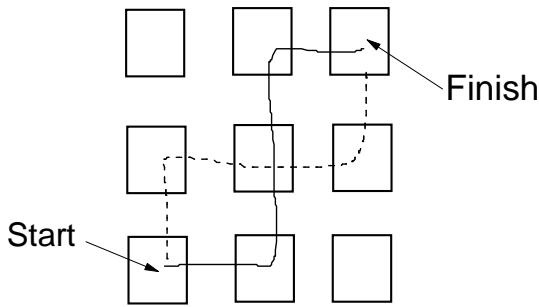}
\vspace{-4pc}
\caption{Adaptive routing of the torus network.  A packet could take
different routes to reach its destination.}
\label{fig:bglrouting}
\end{figure}
Virtual channels are used to in the torus network to avoid dead-locks
\cite{dally87}.  A token based flow control
mechanism is used to improve throughput \cite{dally92}.
The torus switch is highly pipelined and implements virtual cut-through
routing \cite{kk79}.  In addition,
the torus network supports both adaptive and 
deterministic minimal-path routing \cite{duato93,duato99}.
Figure \ref{fig:bglrouting} illustrates
adaptive routing.  Packets could take different paths to reach their
destination as long as the paths are minimal, i.e., every hop a packet
makes should reduce its distance to the destination.  The actual path
a packet goes through is determined dynamically on each node
depending on local traffic.  All these features combined
optimize both the throughput and latency over the torus network.  

In addition to the point-to-point message passing, we have also
implemented multicast operations in the torus hardware
to allow for one node to broadcast a message to a ``class" of nodes.
This feature has been proven to be very useful for a number of
applications.

All data and token flow-control packets are protected by CRC.  If an error
happens on a link, the bad packet that got received will be automatically
deleted from the network, and the good packet will be retransmitted
over the same link.  The error and retransmission protocol is
handled automatically by the torus hardware.

\subsection{Global Tree Network}
On a large scale parallel machine like BlueGene/L, one constraint
that affects the scalability of a large class of applications is the
time it takes for a global reduction operation, for example, a global sum.

\begin{figure}[htb]
\includegraphics[angle=270,scale=0.5,width=\hsize]{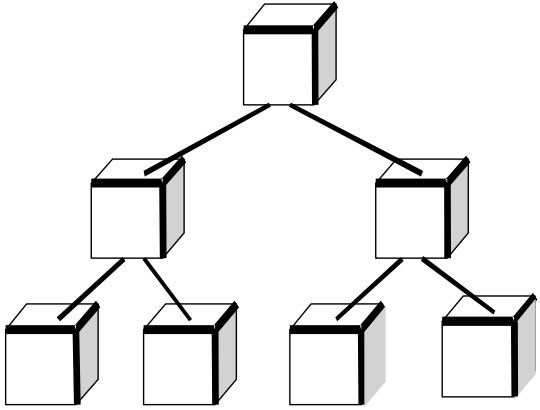}
\vspace{-1pc}
\caption{Global tree network.}
\label{fig:bgltree}
\end{figure}
To reduce the latency of global reduction operations, the BG/L
supercomputer implements a global tree network.  This is a binary-tree
like network shown in Figure \ref{fig:bgltree}.  The global tree
network supports global broadcast from a single node to all nodes, and
global reduction operations including global integer maximum,
global integer sum and certain binary operations like global AND,
OR and XOR.  The tree network logic is also integrated into the node
ASIC.

During a global reduction, each node contributes a message.  All messages
from all nodes are combined into a single message and are
then broadcast to each contributing nodes.  A single round-trip
latency on the global tree network is about 5 ${\mu}s$.  
A global floating point sum (implemented by first an integer max to scale 
the exponents, then an integer sum) requires 2 round-trips over the
network and therefore has a latency of approximately 10 ${\mu}s$.
Each tree link maintains a constant bandwidth of 2.8 Gbit/s per link
per direction.

In addition to the global operations mentioned above, the tree network
also supports point-to-point messaging from the root of a segment of
the tree to all its sub-leaves.  I/O nodes are connected to compute nodes
through the tree network.

\subsection{I/O}
Each BG/L node ASIC has an integrated Gbit Ethernet.  On a compute node,
this Ethernet port is not used.  While on an I/O node, the Gbit Ethernet is
connected to central Gbit Ethernet switches, which in turn connects
to the host computer and I/O disk farm.

Each I/O node connects to a number of compute nodes through the high
speed tree network.  Depending on the I/O requirement, the maximum I/O node
to compute node ratio is one I/O node for every 8 compute nodes.
On BG/L, we plan to install one I/O node for every 64 compute nodes leading
to a total of 1024 I/O nodes for the 64K node machine.  The total peak
I/O bandwidth is therefore 1 Terabit/s.

\subsection{System Partitioning}
On a large scale parallel supercomputer, it is often desirable that a big
machine can be partitioned into smaller machines.  On BG/L, partitioning
is achieved through software configuration.

\begin{figure}[hbt]
\vspace{-1pc}
\includegraphics[angle=270,scale=0.5,width=\hsize]{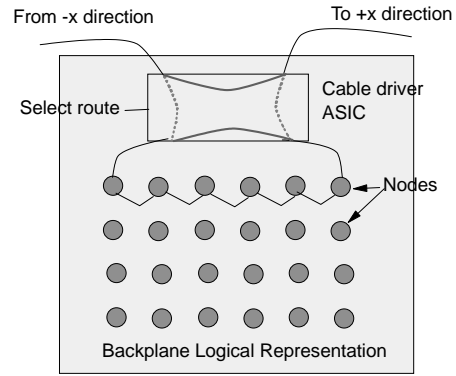}
\vspace{-2pc}
\caption{Illustration of partitioning of BG/L.  A cabinet of nodes can either
be connected in a torus loop, or be jumped over where the cabinet will
form its own torus in a separate physical partition.}
\label{fig:bglpartition}
\end{figure}
Figure \ref{fig:bglpartition} illustrates the basic idea of BG/L
partitioning.  Within a BG/L cabinet, there are 2 mid-planes, each
with 512 compute nodes attached forming a $8\times8\times8$ 3-d lattice.
On the edge of each mid-plane, there is a set of re-drive chips.
They capture the high speed torus and tree network signals coming from node
ASICs over a certain length of mid-plane board trace, then re-drive
them over the cable connecting different cabinets.  
This improves the high speed signal quality. 
These re-drive chips can be programmed by the host to either
include the nodes of the current cabinet in the torus loop for a
large partition, or to skip over them, therefore creating a separate
physical partition.

This scheme allows each partition to be electrically isolated.
Each partition has its own torus and tree networks that
do not communicate with other partitions.  With a few more spare
cabinets in the system and proper diagnostic software, the 
BG/L system can detect failures and swap the cabinet that has bad nodes
with a good cabinet, and automatically restart the application
from the last checkpoint on a complete good machine partition.
This provides added reliability for the BG/L system.

\begin{figure}[hbt]
\vspace{-1pc}
\includegraphics[angle=270,scale=0.5,width=\hsize]{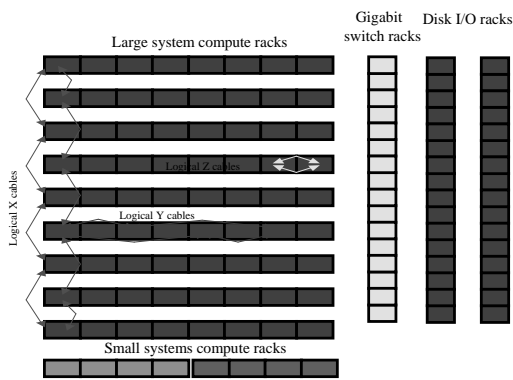}
\vspace{-2pc}
\caption{BlueGene/L system expected to be built at the LLNL.}
\label{fig:bglsystem}
\end{figure}

Figure \ref{fig:bglsystem} shows the complete system expected to be
built at the LLNL.  There are 64 main cabinets plus 8 cabinets for spares
and code development.  These cabinets are then connected through racks
of Gigabit Ethernet switches to disk I/O racks.  A host computer
is connected directly to the Gigabit switch complex.  An example of
partitioning would be to have a large 32 cabinet system
(32$\times$32$\times$32 torus), a number of medium 4 cabinet systems
(16$\times$16$\times$16 torus) and a number of small 1/2 cabinet systems
(8$\times$8$\times$8 torus).

\section{SOFTWARE}
Scalable system software that supports efficient execution of parallel
applications is an integral part of the BlueGene/L architecture.
Our plan is to have a lightweight high-performance kernel running on 
compute nodes, and to expect Linux running on I/O nodes.  The 
lightweight compute kernel approach was motivated by the Puma and
Cougar kernels at Sandia National Laboratory and the University
of New Mexico. 

The BG/L compute kernel is a single user OS
that supports execution of a single dual-threaded (one thread for each 
of the two processors within a node) user process.  It provides a single
and static virtual address space to the one running compute process.
There is no need for context switching.  Because the PowerPC 440
processor core supports large pages, demand paging is not necessary.
The user process will receive full resource utilization,
yet the OS is still protected from the user application through
a virtual memory system.

I/O nodes will support multiple processes.  They will only 
execute system software, no user applications.  They provide I/O
support to compute nodes, i.e., file operations to the disk farm,
control and monitoring support for the host, etc.

In terms of development tools and environment, IBM's XL-C, XL-C++ and
XL-FORTRAN compilers are expected to be ported to support both 
the PPC 440 integer unit as well as the FP2 floating point unit.
The GNU-C compiler already
supports the common set of PowerPC instructions, albeit it does
not support the FP2 instruction extensions, which the IBM compilers
are expected to support.  A set of highly optimized math libraries is also
expected to be provided to facilitate high performance application development.
As for parallel environment, we anticipate to provide MPI support as well
as a user level, low latency system programming interface to facilitate
the utilization of the high speed torus and tree networks.

\section{APPLICATIONS}
The BG/L team has been doing extensive studies of a wide variety of
applications \cite{BGLISSCC02}.  The results of these studies are
very encouraging.  They show that a large class of applications
scales well on the BG/L architecture, even to the level of
65,536 nodes.

In this section, for the interest of the lattice community, we will
compare the architecture of BG/L to QCDOC, as applied to QCD type
applications.  Reference \cite{qcdoc02} shows that for
Wilson fermions, on a $2^4$ lattice per node, QCDOC can sustain
about 50\% of the peak performance.

\begin{table}[htb]
\caption{Comparisons between QCDOC and BlueGene/L.}
\label{tab:bglqcdoc}
\begin{tabular}{lcc}
\hline
 & QCDOC & BlueGene/L \\
\hline
CMOS Technology	& 0.18 $\mu m$ 		& 0.13 $\mu m$ \\
CPU		& 500 MHz		& 700 MHz$\times$2	\\
EDRAM size	& 4 MB			& 4 MB			\\
Peak flops/node	& 1 Gflops		& 5.6 Gflops		\\
Network Topology& 6-d torus		& 3-d torus		\\
\hline
Bandwidth:	&			&			\\
L1 cache to FPU	& 4 GB/s		& 22.4 GB/s		\\
CPU interface peak & 8 GB/s		& 11.2 GB/s		\\
CPU read sustained & 3 GB/s		& 8.4 GB/s		\\
EDRAM interface	& 16 GB/s		& 22.4 GB/s		\\
External DRAM	& 2.6 GB/s		& 5.6 GB/s		\\
Torus/link	& 0.5 Gbit/s		& 1.4 Gbit/s		\\
\hline
\end{tabular}
\end{table}
Table \ref{tab:bglqcdoc} shows a comparison of various performance
and memory bandwidth numbers between QCDOC and BG/L. 
For Lattice QCD applications, because
most of the communications are between nearest neighbors,
we expect that both processors on a BG/L node will be available 
for computation.  From QCDOC to BG/L on a per node basis,
the sustained memory read
bandwidth from EDRAM improves by a factor of 2.8, the external
memory bandwidth by a factor of 2.15 and the torus network bandwidth
by a factor of 2.8.  While the peak floating performance is increased
from 1 Gflops to 5.6 Gflops, we expect the overall performance
improvement per node will be determined mainly by the memory bandwidth, 
therefore about a factor 2.15 to 2.8.  Given the high efficiency of
QCDOC, we expect BG/L to also perform well for QCD applications.

\section{CONCLUSION}
System-on-a-chip technology has opened new possibilities of building
large scale supercomputers.  By optimizing the overall system, great
reduction in cost/performance, power consumption/performance and
machine size/performance can be achieved.

BlueGene/L is a first step in IBM's commitment to petaflops
scale computing by exploring a new architecture of building massively
parallel machines.  With more applications ported and
optimized for this kind of architecture, there will be great
scientific benefits and rewards.

\end{document}